\documentclass[12pt,aps,showpacs,eqsecnum,nofootinbib,floatfix]{revtex4}
\usepackage{bbding}
\usepackage{pifont}
\usepackage{subfigure}
\usepackage{epsfig}
\usepackage{amsmath}
\usepackage{amsfonts}
\usepackage{amssymb}
\usepackage{color}
\usepackage{graphicx}
\usepackage{mathrsfs}
\usepackage{multirow}

\def\be{\begin{equation}}
\def\ee{\end{equation}}
\def\ba{\begin{eqnarray}}
\def\ea{\end{eqnarray}}

\newcommand{\omits}[1]{}

\begin{document}
%\begin{CJK*}{GBK}{song}

\title{Thermodynamics and critical behaviors of topological dS black holes
with nonlinear source}
\author{Hui-Hua Zhao, Li-Chun Zhang, Fang Liu, Ren Zhao\footnote{corresponding author: Email: kietemap@126.com(Ren Zhao)}}

\medskip
\affiliation{Institute of Theoretical Physics, Shanxi Datong
University, Datong 037009, China}
\affiliation{Department of
Physics, Shanxi Datong University, Datong 037009, China}

\begin{abstract}

We discuss black hole solutions of Einstein gravity
in presence of nonlinear electrodynamics in dS spacetime.
Considering prescribed entropy, thermodynamic volume of dS spacetime,
We investigate properties of the effective thermodynamic quantities under influence of
nonlinearity parameter $\alpha$. They show a similar phase transition and criticality
properties with that of black holes in AdS spacetime. And the nonlinearity parameter
$\alpha$ combined with electric charge is found to have effects on the phase structure.
By the Ehrenfest equations we prove the critical phase transition is a second order
equilibrium transition.

\textbf{Keywords}: de Sitter spacetime; phase transition; nonlinear source;
effective thermodynamic quantities; stability
\end{abstract}
\pacs{04.70.-s, 05.70.Ce}

\maketitle

\bigskip

\section{Introduction}

Since the four laws of black hole mechanics was discovered, black
holes are widely believed to be thermodynamic and they possess
standard thermodynamic quantities such as temperature and
entropy\cite{Bekenstein1973, Bardeen1973}. The particular
thermodynamic quantities of black hole and its holographic
property are quantum essentially, so black hole is a 
macroscopical quantum system. Therefore, the studies of black hole
thermodynamic properties provide an important window to investigate
quantum gravity\cite{Bekenstein1973, Bardeen1973, Hawking1974}.
Phase transition is one of interesting subjects in
studying a thermodynamic system. It is found that black holes may go through some
interesting phase transition, for example the Hawking-Page phase
transition \cite{Hawking1983}, and have some critical phenomena similar to that of
usual thermodynamic systems. 

Theoretically the cosmological constant term is expected to arise
from the vacuum expectation value of a quantum field and hence can
vary. Therefore, it may be considered in the first law of
thermodynamics with its conjugate \cite{Gibbons1996,Brown1987,Caldarelli2000}. By this
generalization, the cosmological constant $\Lambda $ and its
conjugate can be interpreted as thermodynamic pressure and volume of
a black object system respectively. For d-dimensional spacetime the
thermodynamic pressure $P$ and its conjugate thermodynamic volume $V$
are defined as\cite{Kubiznak2012,Dolan2013,Gunasekaran2012,Cvetic2011}
\begin{equation}
\label{eq1}
P=-\frac{1}{8\pi }\Lambda =\frac{(d-1)(d-2)}{16\pi l^2}
\end{equation}
\begin{equation}
\label{eq2}
V=\left( {\frac{\partial M}{\partial P}} \right)_{S,Q_i ,J_k } .
\end{equation}
Recently critical behaviors and phase transitions of black holes
have been extensively investigated by considering the cosmological
constant as thermodynamic pressure
\cite{Kastor2011,Banerjee12011,Banerjee22011,Banerjee32011,
Cai12013,Ma12014,Zhao12013,zhao22013,Ma22014,Ma42014,Zou12014,
Zou22014,Wei2013,Wei2015,Zangeneh2016,Liu2014,Zhaohh2015}. It is interesting that the
studies on the charged black holes show they may have an analogous
phase transition with that of van der Waals-Maxwell's liquid-gas.
People have also been trying to construct a complete liquid-gas
analogue system for black holes\cite{Hendi2017,Dehyadegaria2017}.

Nonlinear field theories are of interest to different branches of
mathematical physics because most physical systems are inherently nonlinear
in the nature. The nonlinear electrodynamics (NLED) theories are
considerably richer than the Maxwell field and in special case they reduce
to the linear Maxwell theory. Recently, It was shown that NLED objects can
remove both the big bang and black hole singularities\cite{Ayon-Beato11999,Ayon-Beato21999,Lorenci2002,Dymnikova2004,Corda2010,Corda2011}.
The first attempt to couple the NLED with gravity was made by Hoffmann\cite{Hoffmann1935}. After that the
effects of Born-Infeld (BI) NLED coupled to the gravitational field
have been extensively studied. In the paper of Hendi \cite{Hendi2013,Hendi2015,Hendi2016,Hendi12017}, thermodynamic
properties of AdS black holes with nonlinear source have been
discussed. We take into account a NLED source and investigate the
effects of nonlinearity on the phase transition properties  of black
holes in dS spacetime in this paper.

In the era of inflation, the Universe is in a quasi-dS space.
The cosmological constant corresponds to vacuum energy and is
usually considered as a candidate for dark energy. The accelerating
Universe will evolve into another dS phase. In order to
construct the entire evolution history of our Universe, we should
have a clear perspective on the classical and quantum properties of
dS space. It is known that with appropriate parameters dS spacetimes possess not only
black hole horizon but cosmological horizon. Moreover both horizons have
thermal radiation but they are of different temperatures, and the two sets
of thermodynamic quantities for the both horizons respectively satisfy the first
law of thermodynamics \cite{Cai22002,Sekiwa2006}. Recently, the study on the physical
properties of dS spacetime has aroused great interest\cite{Akhmedov2014,Chen2012,Arraut2013,Zhao32014}.
Take into account the thermodynamic relevance of the two horizons of dS
spacetime and analyze the first law of thermodynamics
satisfied by the two horizons, the Ref\cite{Urano2009,Ma52014,Zhaohh2014,Ma62015,Guo2015} obtained the effective
temperature, effective pressure and effective electric potential.
Moreover the thermodynamic behaviors of some dS spacetimes were discussed
by the effective thermodynamics quantities. In this paper,
we discuss the critical properties of charged dS spacetime with nonlinear source(RN-dS-N system).
Based on the effective thermodynamic quantities of the RN-dS-N system,
we put emphasis on discussing the effects of nonlinear electrodynamics
disturbance on the critical behaviors of dS spacetime.

Outline of the paper is as follow: In the next section, we introduce the
topological black holes with nonlinear source and the two sets of
thermodynamic quantities corresponding to the two horizons of dS spacetime.
In the third section, we analyze the relations of the thermodynamic
quantities and consider the dS spacetime as a whole, then get the effective
thermodynamic quantities. The critical behaviors of the dS
spacetime as a thermodynamic system and the effect the nonlinearity parameters on
critical behaviors are analysed in the fourth section. And in fifth section, by Ehrenfest
scheme we testify the critical behaviors belong to the second order phase transition.
In the last section, we make some discussions and conclusions.
(We use the units $G_d =\hbar =k_B =c=1$)

\section{Topological black holes with nonlinear source}

The $(n+1)$-dimensional action of Einstein gravity of nonlinear
electrodynamics is\cite{Hendi12015,Hendi22015}:%[108,109]
\begin{equation}
\label{eq3}
I_G =-\frac{1}{16\pi }\int_M {d^{n+1}x\sqrt {-g} } [R-2\Lambda
+L(F)]-\frac{1}{8\pi }\int_{\partial M} {d^nx\sqrt {-\gamma } \Theta (\gamma
)} ,
\end{equation}
where $R$ is the scalar curvature, $\Lambda $ is the cosmological constant.
In this action,
\begin{equation}
\label{eq4}
L(F)=-F+\alpha F^2+O(\alpha ^2),
\end{equation}
is the Lagrangian of nonlinear electrodynamics. $F=F_{\mu \nu } F^{\mu \nu }$ is the Maxwell invariant, in which
$F_{\mu \nu } =\partial _\mu A_\nu -\partial _\nu A_\mu $ is the
electromagnetic field tensor and $A_\mu $ is the gauge potential. In
addition, $\alpha $ denotes nonlinearity parameter which is small, so the
effects of nonlinearity should be considered as a perturbation.

The $(n+1)$-dimensional topological black hole solutions can take the form
of \cite{Arraut2013}
\begin{equation}
\label{eq5}
ds^2=-f(r)dt^2+\frac{dr^2}{f(r)}+r^2d\Omega _{n-1}^2 ,
\end{equation}
where

\begin{equation}
\label{eq6} f(r)=k-\frac{m}{r^{n-2}}-\frac{2\Lambda
r^2}{n(n-1)}+\frac{2q^2}{(n-1)(n-2)r^{2n-4}}-\frac{4q^4\alpha
}{[2(n-2)(n+2)+(n-3)(n-4)]r^{4n-6}}.
\end{equation}
$m$ is an integration constant which is related to the mass of
the black hole and the last term in Eq. (\ref{eq6}) indicates the
effect of nonlinearity. The asymptotical behavior of the solution is
AdS or dS provided $\Lambda <0$ or $\Lambda >0$ and the case of
asymptotically flat solution is permitted for $\Lambda =0$ and
$k=1$.

When $\Lambda >0$, the black holes have black hole horizon and cosmological
horizon, and $f(r_{+,c} )=0$. The temperatures at the two horizons
respectively are \cite{Hendi12015}%[108]
\begin{equation}
\label{eq7}
T_+ =\frac{f'(r_+ )}{4\pi }=\frac{1}{2\pi (n-1)}\left(
{\frac{(n-1)(n-2)k}{2r_+ }-\Lambda r_+ -\frac{q^2}{r_+^{2n-3}
}+\frac{2q^4\alpha }{r_+^{4n-5} }} \right),
\end{equation}
\begin{equation}
\label{eq8} T_c =-\frac{f'(r_c )}{4\pi }=-\frac{1}{2\pi (n-1)}\left(
{\frac{(n-1)(n-2)k}{2r_c }-\Lambda r_c -\frac{q^2}{r_c^{2n-3}
}+\frac{2q^4}{r_c^{4n-5} }} \right).
\end{equation}
The ADM (Arnowitt-Deser-Misner) mass and electric charge parameter $Q$ per unit volume $V_{n-1} $ of
the black hole are

\begin{equation}
\label{eq9+}
M=\frac{V_{n-1} \left( {n-1} \right)m}{16\pi },
\end{equation}
\begin{equation}
\label{eq9}
Q=\frac{q}{4\pi }V_{n-1} ,
\end{equation}
with $V_{n-1} =\frac{2\pi ^{n \mathord{\left/ {\vphantom {n 2}} \right.
\kern-\nulldelimiterspace} 2}}{\Gamma \left( {n \mathord{\left/ {\vphantom
{n 2}} \right. \kern-\nulldelimiterspace} 2} \right)}$.

The mass of the black hole can be expressed as
\[
 M=\frac{V_{n-1} (n-1)}{16\pi }\left( kr_{+,c}^{n-2}
-\frac{2\Lambda r_{+,c}^n}
{n(n-1)}+\frac{2q^2}{(n-1)(n-2)r_{+,c}^{n-2} } \right.
\]

\begin{equation}
\label{eq10}
\left. -\frac{4q^4\alpha
}{[2(n-2)(n+2)+(n-3)(n-4)]r_{+,c}^{3n-4} } \right)
\end{equation}

The entropy and thermodynamic volume of the black hole corresponding to
black hole horizon and cosmology horizon are\cite{Dolan2013}

\begin{equation}
\label{eq11}
 S_+ =\frac{V_{n-1} r_+^{n-1} }{4}, S_c =\frac{V_{n-1}
r_c^{n-1} }{4}, \quad V_+ =\frac{V_{n-1} r_+^n }{n}, \quad V_c
=\frac{V_{n-1} r_c^n }{n}.
\end{equation}

Recently, in the view of that cosmological constant is considered as
thermodynamic pressure of black holes\cite{Kubiznak2012,Dolan2013,Gunasekaran2012,Cvetic2011}, substituting
eq.(\ref{eq11}) into eqs.(\ref{eq7}) and (\ref{eq8}) two
thermodynamic systems can be obtained corresponding to black hole
horizon and cosmological horizon respectively. On account of the two
thermodynamic systems are independent of each other, their
thermodynamic properties have been researched and achieved some
meaningful results\cite{Dolan2013,Sekiwa2006}.

\section{Effective thermodynamic quantities of dS spacetime}

The thermodynamics volume and entropy of
spherically symmetric dS spacetime satisfy\cite{Dolan2013,Sekiwa2006}
\begin{equation}
\label{eq12} V=V_c -V_+ =\frac{V_{n-1} r_c^n }{n}\left( {1-x^n}
\right), \quad S=S_c +S_+ =\frac{V_{n-1} r_c^{n-1} }{4}\left(
{1+x^{n-1}} \right).
\end{equation}
in which, $x=\frac{r_+ }{r_c }\le 1$. From Eq. (\ref{eq10}), one can
see that for the dS spacetime with black hole horizon and
cosmological horizon both the positions of the two horizons $r_+ $
and $r_c $ are the functions of the spacetime energy (mass)$M$,
electric charge$Q$, and cosmological constant$\Lambda$. So $r_+ $ and $r_c $ are
not independent of each other. Therefore if the
two horizons are considered as two thermodynamic systems, they are
not independent. We should take into account the
relevance in studying the thermodynamic properties of dS spacetime.

Using the Eqs.(\ref{eq12}) and (\ref{eq10}), the energy (mass)$M$
can be expressed as
\[
M=\frac{V_{n-1} (n-1)}{16\pi (1-x^n)}r_c^{n-2} \left[ {k(x^{n-2}-x^n)}
\right.
+\frac{2q^2}{(n-1)(n-2)r_c^{2(n-2)} }\frac{(1-x^{2n-2})}{x^{n-2}}
\]
\begin{equation}
\label{eq13} \left. {-\frac{4q^4\alpha
}{[2(n-2)(n+2)+(n-3)(n-4)]r_c^{2(2n-3)}
}\frac{(1-x^{4n-4})}{x^{3n-4}}} \right].
\end{equation}
Combining Eqs. (\ref{eq12}) and (\ref{eq13}), as $\alpha $ is a
constant, one can see that the energy (mass) of the system is a function
of entropy, thermodynamic volume, and electric charge, that is
\begin{equation}
\label{eq14} M=M(S,V,Q).
\end{equation}
It is well known that when the energy of the spacetime acts as a
function of the thermodynamic quantities corresponding to black hole
horizon or those corresponding to cosmological horizon, the
functions meet the first law of thermodynamics \cite{Dolan2013}%[19]
\begin{equation}
\label{eq15} \delta M=T_+ \delta S_+ +\Phi _+ \delta Q+V_+ \delta
P_0 , \quad \delta M=-T_c \delta S_c +\Phi _c \delta Q+V_c P_0 ,
\end{equation}
in which $\Phi _+$ and $\Phi _c$ are the charge potential at black hole
horizon and at cosmological horizon respectively and
\begin{equation}
\label{eq16} P_0 =-\frac{\Lambda }{8\pi }.
\end{equation}
There exist two baffling problems considering the two thermodynamic
systems presented in the Eq.(\ref{eq15}): Firstly, the energy
$M$, electric charge $Q$ and cosmological constant $\Lambda$ are the
common state parameters of the two systems. Thus the thermodynamic
quantities of the two systems are dependent. Secondly, the black
hole horizon radiation temperature $T_+$ and the cosmological
radiation temperature $T_c$ present in Eq.(\ref{eq15}) are usually
different. So dS spacetime is in nonequilibrium. At
present, a mature theory has not been found to analyze the
nonequilibrium thermodynamic system. The characteristic that the two thermodynamic systems own the common state
parameters reminds us to build an effective thermodynamic system to reflect
the thermodynamics properties of the dS spacetime.
\begin{equation}
\label{eq17} dM=TdS+\Phi dQ-PdV.
\end{equation}
The effective temperature $T$, effective electric potential $\Phi $,
and effective pressure $P$ respectively are
\begin{equation}
\label{eq18}
T=\left( {\frac{\partial M}{\partial S}} \right)_{Q,V} =\frac{\left(
{\frac{\partial M}{\partial x}} \right)_{r_c,q } \left(
{\frac{\partial V}{\partial r_c }} \right)_{x,q} -\left(
{\frac{\partial V}{\partial x}} \right)_{r_c,q } \left(
{\frac{\partial M}{\partial r_c }} \right)_{x,q}}{\left(
{\frac{\partial S}{\partial x}} \right)_{r_c,q} \left(
{\frac{\partial V}{\partial r_c }} \right)_{x,q} -\left(
{\frac{\partial V}{\partial x}} \right)_{r_c,q} \left(
{\frac{\partial S}{\partial r_c }} \right)_{x,q} },
\end{equation}
%\[
%=\frac{1}{4\pi x^{n-2}(1-x^n)(1+x)r_c }
%\]
%\[
%\left[ k\left(
%x^{n-3}(n-2-nx^2)(1-x^n)+nx^{n-1}(x^{n-2}-x^n)+(n-2)x^{n-1}(x^{n-2}-x^n)
%\right) \right.
%\]
%\[
%+\frac{2q^2}{(n-1)(n-2)r_c^{2(n-2)} }\left(
%{\frac{(2n-2)x^n-nx^{2n-2}-(n-2)}{x^{n-1}}-\frac{(n-2)(1-x^{2n-2})}{x^{-1}}}
%\right)
%\]
%\[
%+\frac{4q^4\alpha }{[2(n-2)(n+2)+(n-3)(n-4)]r_c^{4n-6} }
%\]
%\begin{equation}
%\label{eq18} \left. {\left(
%{(3n-4)\frac{(1-x^{4n-4})}{x^{2n-3}}-\frac{(4n-4)x^n-nx^{4n-4}-(3n-4)}{x^{3n-3}}}
%\right)} \right],
%\end{equation}
\begin{equation}
\label{eq19}
\Phi =\left( {\frac{\partial M}{\partial Q}} \right)_{S,V} =\left(
{\frac{\partial M}{\partial q}\frac{\partial q}{\partial Q}}
\right)=\frac{(n-1)(1-x^{2n-2})q}{(1-x^n)r_c^{n-2} x^{n-2}}\left[
{\frac{1}{(n-1)(n-2)}} \right.
\end{equation}
%\begin{equation}
%\label{eq19} \left. {-\frac{4q^2\alpha
%}{[2(n-2)(n+2)+(n-3)(n-4)]r_c^{2n-2} }\frac{(1+x^{2n-2})}{x^{2n-2}}}
%\right],
%\end{equation}
\begin{equation}
\label{eq20}
P=-\left( {\frac{\partial M}{\partial V}} \right)_{Q,S} =-\frac{\left(
{\frac{\partial M}{\partial x}} \right)_{r_c } \left( {\frac{\partial
S}{\partial r_c }} \right)_x -\left( {\frac{\partial S}{\partial x}}
\right)_{r_c } \left( {\frac{\partial M}{\partial r_c }} \right)_x }{\left(
{\frac{\partial V}{\partial x}} \right)_{r_c } \left( {\frac{\partial
S}{\partial r_c }} \right)_x -\left( {\frac{\partial S}{\partial x}}
\right)_{r_c } \left( {\frac{\partial V}{\partial r_c }} \right)_x }
\end{equation}

%\[
%=\frac{1}{16\pi (1+x)(1-x^n)^2}\left[ {\frac{-\left( {n-1} \right)k}{r_c^2
%x^2}\left( {\left( {n-2} \right)\left( {x^{2n+2}-x}
%\right)-nx^{2n}+2x^{n+2}-2x^{n+1}+nx^3} \right)} \right.
%\[
%+\frac{2q^2\left( {x^n+x} \right)}{r_c^{2n-2} x^{2n}}\left(
%{\frac{(2n-2)\left( {x^{n+2}-x^{2n}} \right)}{(n-2)}+\left( {x^n-x}
%\right)\left( {x^{2n}+x} \right)} \right)
%\]

%\begin{equation}
%\label{eq20} -\frac{4q^4\alpha (x+x^n)}{r_c^{4n-4} x^{4n}}\left.
%{\left( {\frac{(4n-4)\left( {x^{n+4}-x^{4n}} \right)}{(3n-4)}\left(
%{x^n-x} \right)\left( {x^{2n}+x} \right)\left( {x^{2n}+x^2} \right)}
%\right)} \right].
%\end{equation}
The Eq.(\ref{eq17}) reflects thermodynamic properties of the whole
dS spacetime rather than that of a horizon, therefore it
gives more comprehensive view of dS spacetime.

\section{Phase transition in topological black holes with nonlinear
source spacetime}

On the basis of the previous section, we study the phase transition and critical behaviors of the
RN-dS-N system. We analyze the effective thermodynamic quantities by Van der
Waals equation, and investigate the relation of the effective pressure and
thermodynamic volume when the temperature is kept constant.
Using the Gibbs free energy criterion, we analyze the phase transition of the system.

When the electric charge $q$ and the nonlinearity parameter $\alpha $ are
kept as constant, the critical point can be obtained by
\begin{equation}
\label{eq21} \left( {\frac{\partial P}{\partial V}} \right)_T =0,
\quad \left( {\frac{\partial ^2P}{\partial V^2}} \right)_T =0.
\end{equation}

\begin{equation}
\label{eq22} \left( {\frac{\partial P}{\partial V}}
\right)_T=\frac{\frac{\partial (P,T)}{\partial (x,r_c
)}}{\frac{\partial (V,T)}{\partial (x,r_c )}}=f(x,r_c )=0,
\end{equation}

\begin{equation}
\label{eq23} \left( {\frac{\partial ^2P}{\partial V^2}} \right)_T
=\left( {\frac{\partial f}{\partial V}} \right)_T
=\frac{\frac{\partial (f,T)}{\partial (x,r_c )}}{\frac{\partial
(V,T)}{\partial (x,r_c )}}=0,
\end{equation}

By the Eqs. (\ref{eq12}), (\ref{eq18}), (\ref{eq20}) and (\ref{eq21})
the critical quantities can be derived when
$Q$, $\alpha$ and dimensionality $n$ are given certain values.  Table 1 below shows some critical
values at dimensionality $n=3$.

\begin{table}[htbp]
\begin{center}
\caption[]{\it Critical values of the effective thermodynamic system
for different $Q$ and $\alpha$ as $n=3$}
\begin{tabular}{|p{23pt}|p{43pt}|p{43pt}|p{43pt}|p{43pt}|p{43pt}|p{43pt}|p{43pt}|p{43pt}|p{43pt}|}
\hline
$n$=3&
\multicolumn{5}{|p{215pt}|}{$Q$=1} &
$Q$=2&
\multicolumn{3}{|p{129pt}|}{$Q$=3}  \\
\hline
&
$\alpha \mbox{=}0.001$&
$\alpha \mbox{=}0.01$&
$\alpha \mbox{=}0.1$&
$\alpha \mbox{=}0.3$&
$\alpha \mbox{=}0.6$&
$\alpha \mbox{=}0.1$&
$\alpha \mbox{=}0.1$&
$\alpha \mbox{=}0.3$&
$\alpha \mbox{=}0.6$ \\
\hline
$x^c$&
0.73222&
0.73230&
0.73316&
0.73559&
0.74274&
0.13443&
0.08697&
0.15771&
0.23708 \\
\hline
$r_c^c $&
3.5058&
3.5025&
3.4675&
3.3760&
3.1627&
7.5636&
13.915&
10.554&
8.8944 \\
\hline
$T^c $&
0.008015&
0.008018&
0.008053&
0.008139&
0.008312&
4.103E-5&
6.541E-6&
4.562E-5&
1.601E-4 \\
\hline
$P^c $&
6.055E-4&
6.061E-4&
6.128E-4&
6.298E-4&
6.666E-4&
1.775E-6&
1.550E-7&
1.409E-6&
5.746E-6 \\
\hline
$V^c$&
109.64&
109.30&
105.82&
97.019&
78.215&
1807.9&
11278.4&
4904.4&
2908.1 \\
\hline $M^c$& 1.2086& 1.2080& 1.2007& 1.1821& 1.1404& 2.1764&
3.6970& 3.1454&
2.9283 \\
\hline $S^c$& 59.315& 59.207& 58.077& 55.179& 48.758& 182.96&
612.89& 358.61&
262.50 \\
\hline $G^c$& 0.7996& 0.7995& 0.7979& 0.7942& 0.7873& 2.172& 3.695&
3.136&
2.903 \\
\hline
\end{tabular}
\label{tab1}
\end{center}
\end{table}

Under the conditions $0<x<1$ and $T>0$,
we derive the relation of $P$ and $V$ numerically as $T$ take some certain values nearby
the critical temperature $T^c$. And we depict these effective isotherms of the RN-dS-N system in the Fig.1. 

\begin{figure}[!htbp]
\center{\subfigure[~$Q=1$,$\alpha=0.1$] {
\includegraphics[angle=0,width=4cm,keepaspectratio]{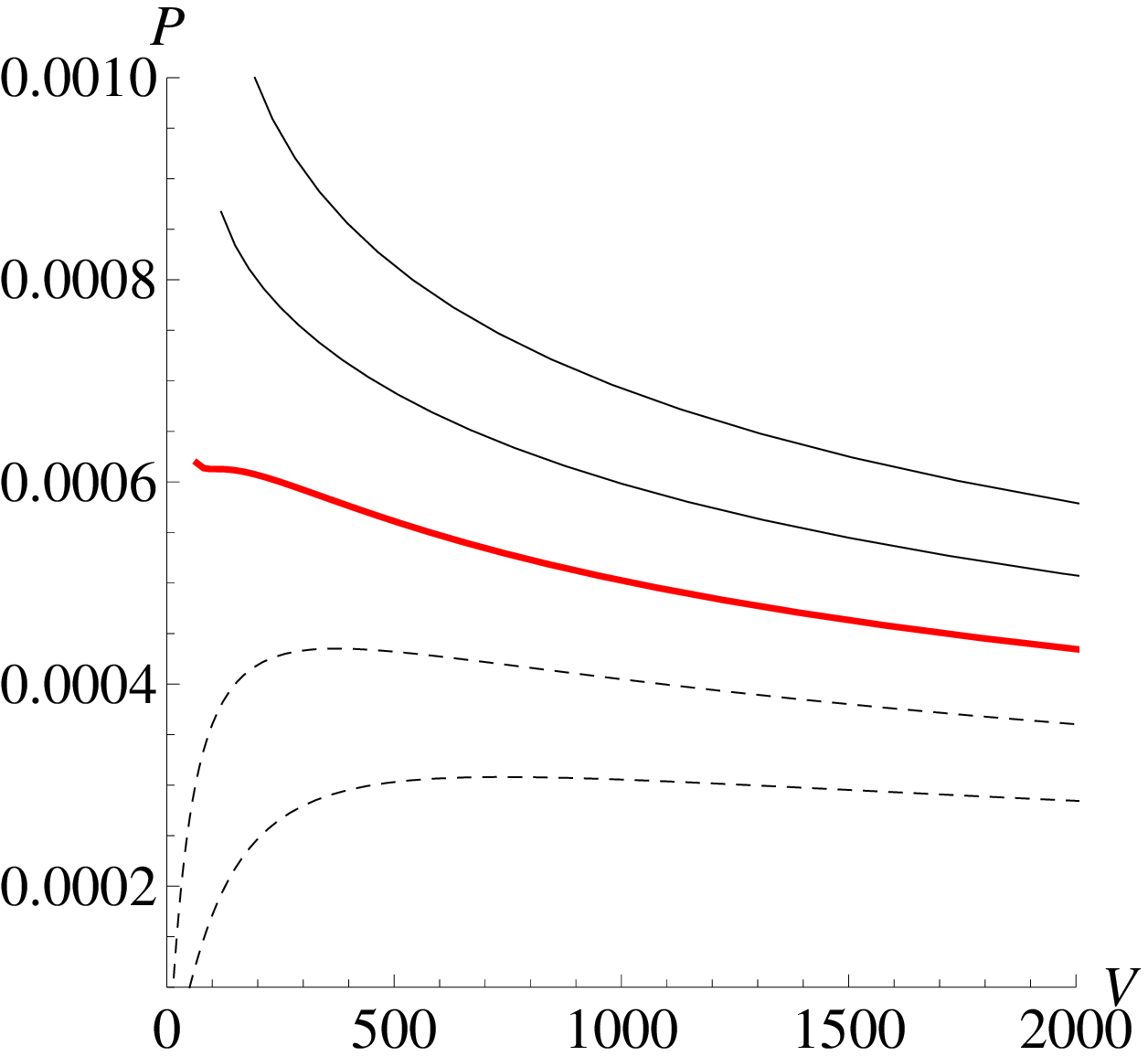}}
\subfigure[~$Q=1$,$\alpha=0.3$] {
\includegraphics[angle=0,width=4cm,keepaspectratio]{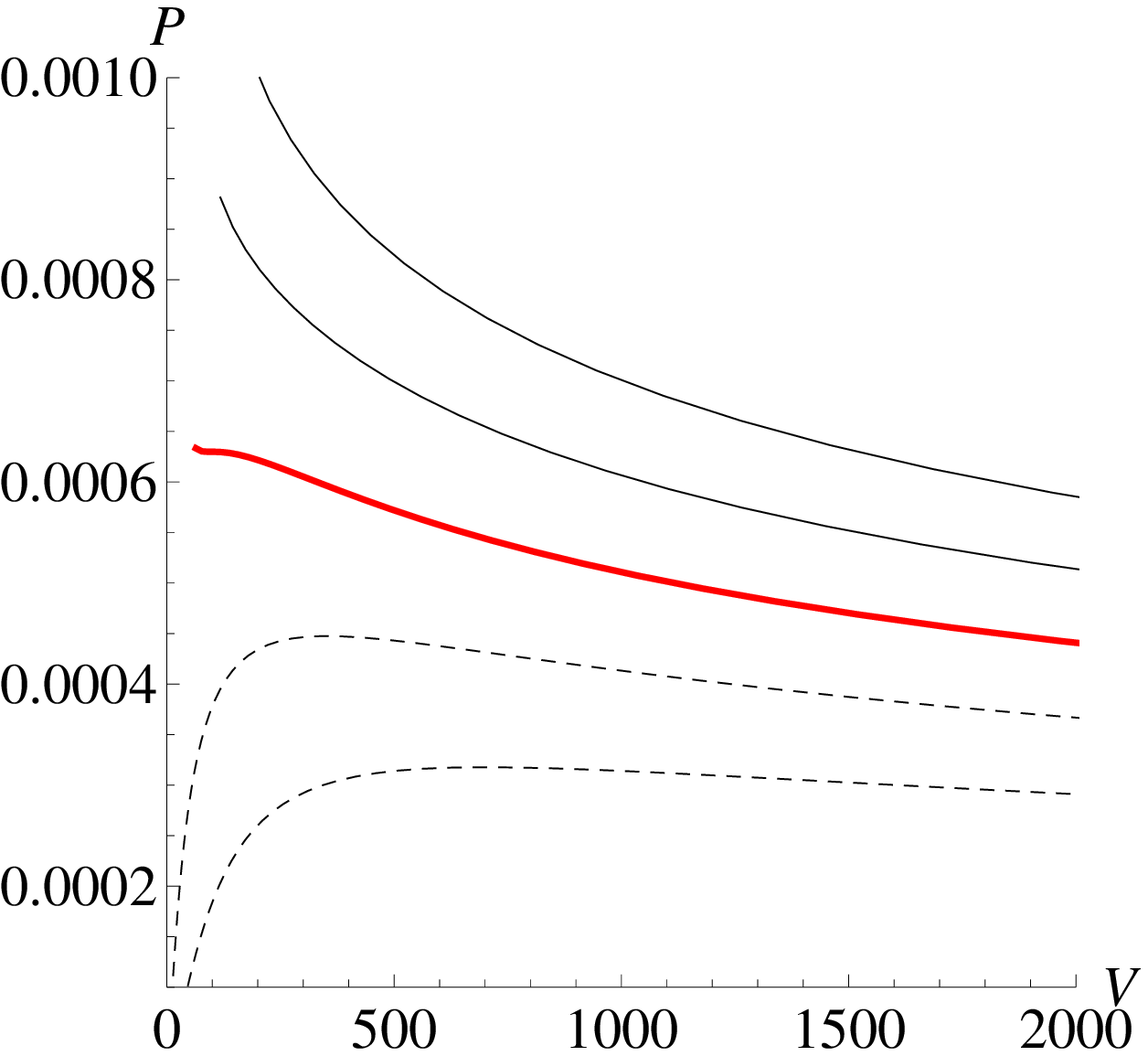}}
\subfigure[~$Q=1$,$\alpha=0.6$] {
\includegraphics[angle=0,width=4cm,keepaspectratio]{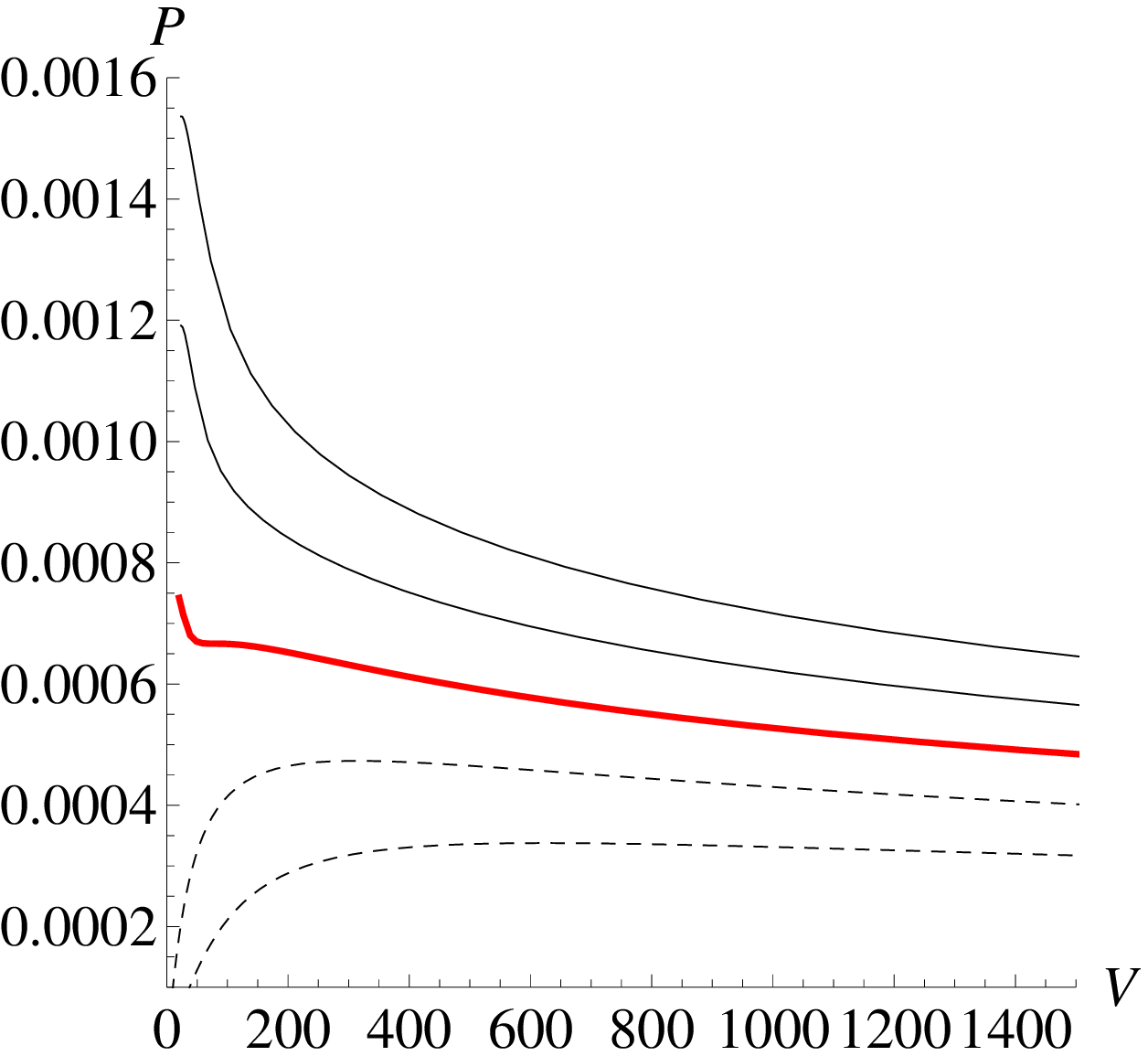}}
\subfigure[~$Q=2$,$\alpha=0.1$] {
\includegraphics[angle=0,width=4cm,keepaspectratio]{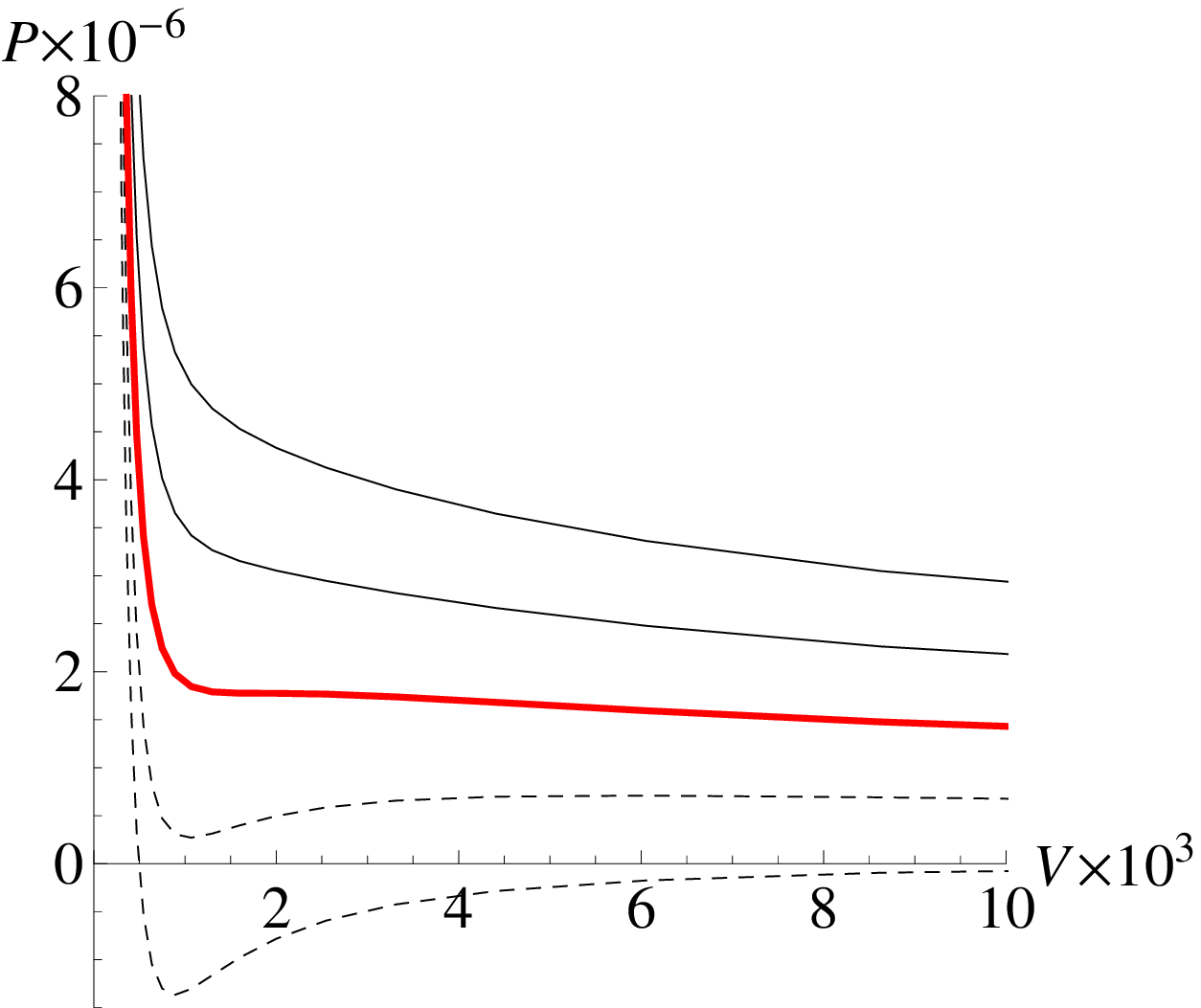}}
\subfigure[~$Q=3$,$\alpha=0.1$] {
\includegraphics[angle=0,width=4cm,keepaspectratio]{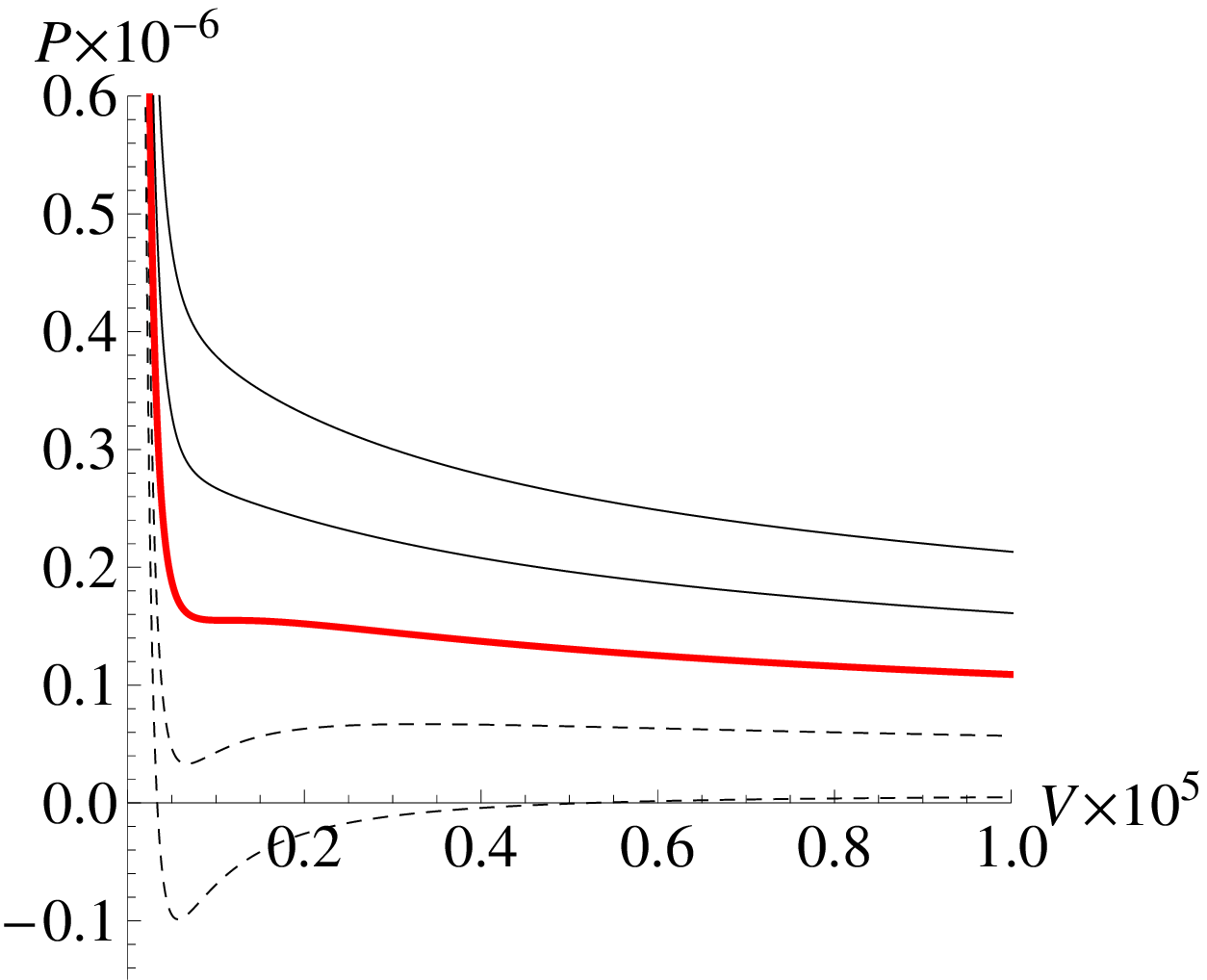}}
\subfigure[~$Q=3$,$\alpha=0.3$] {
\includegraphics[angle=0,width=4cm,keepaspectratio]{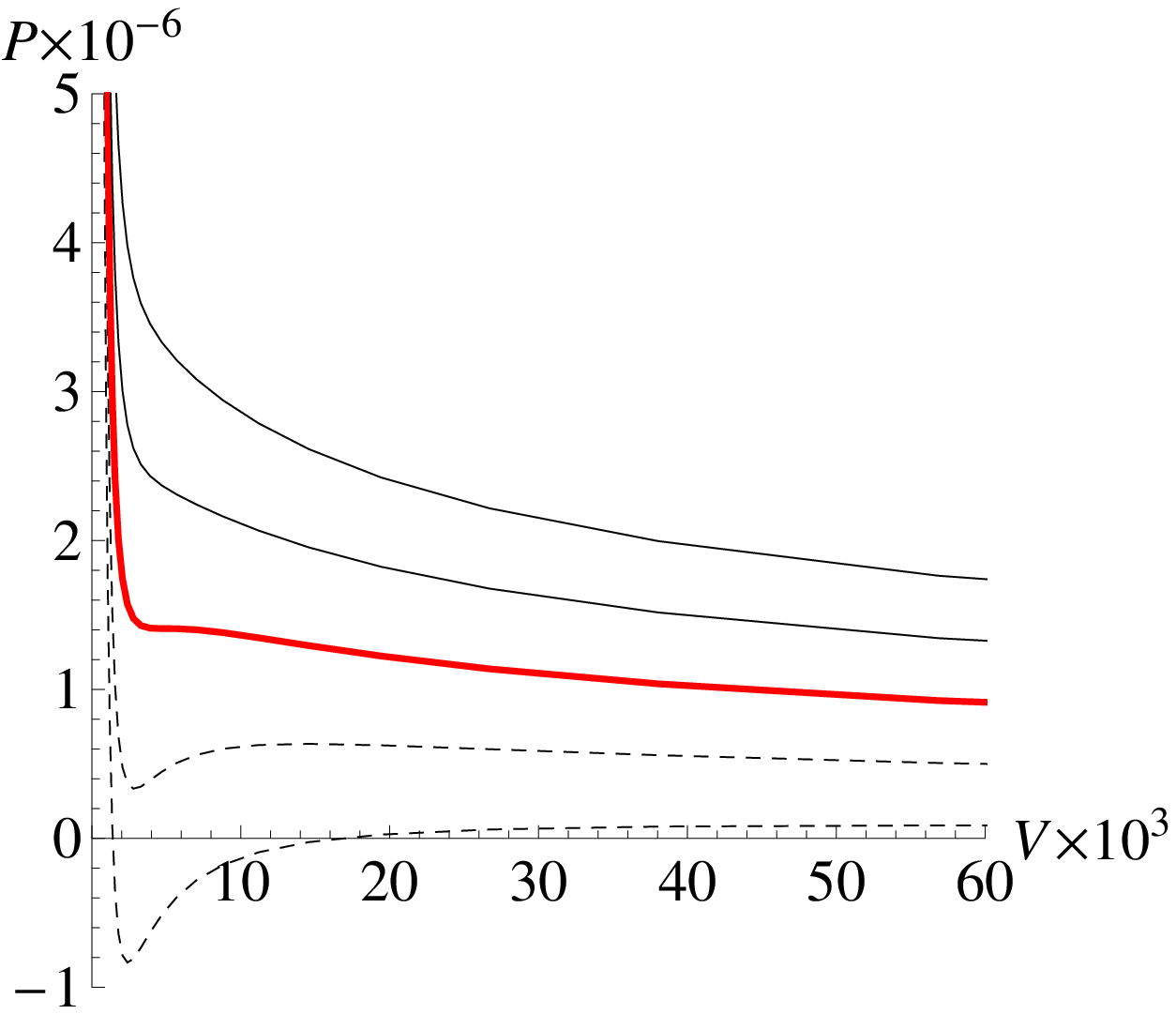}}
\caption[]{\it Effective isotherms in $P-V$ diagrams of $n+1$
dimensional RN-dS-N system as $n=3$. The
dashed lines match $T<T^c$,  the thick solid lines match $T=T^c$, and the thin
solid lines match $T>T^c$.}}
\label{PV}
\end{figure}

It can be seen in Table 1 that with
increasing $\alpha$  all of the critical values $x^c$, the critical temperature $T^c$ and the critical pressure $P^c$
are increasing as $Q$ is given. And when $Q$ is greater the increases are more obvious. 
The other critical quantities in the table go just the opposite. 
Similarly, as $\alpha=0.1$, all of $x^c$, $T^c$ and $P^c$ are increasing with increasing $Q$, 
the other critical quantities in the table go the opposite.
 
Both Table 1 and Fig. 1 show that the nonlinear parameters $\alpha$ and the electric charge $Q$ 
influence the critical state together. Fig.1 shows that as either of the values of  $\alpha$ and $Q$ is greater or 
both of them are greater the more acceptable physical states exist within the smaller volume 
range, that is, the phase structure is more completed and it is more like that in AdS spacetime.
It can not be seen in the Fig.1 whether there exist stable phase transition, especially in Fig.1(a) and Fig.1(b). 
We discuss the phase transition near the effective critical temperature $T^c$ by Gibbs free energy criterion.

Gibbs free energy defined \cite{Mo2014} as
\begin{equation}
\label{eq24} G=M-TS+PV
\end{equation}
Under the conditions $0<x<1$ and $T>0$, we depict the $G-P$ diagrams with some effective isotherms 
nearby the critical temperature $T^c$ for the RN-dS-N system in Fig.2.

\begin{figure}[!htbp]
\center{\subfigure[~$Q=1$,$\alpha=0.1$] {
\includegraphics[angle=0,width=4cm,keepaspectratio]{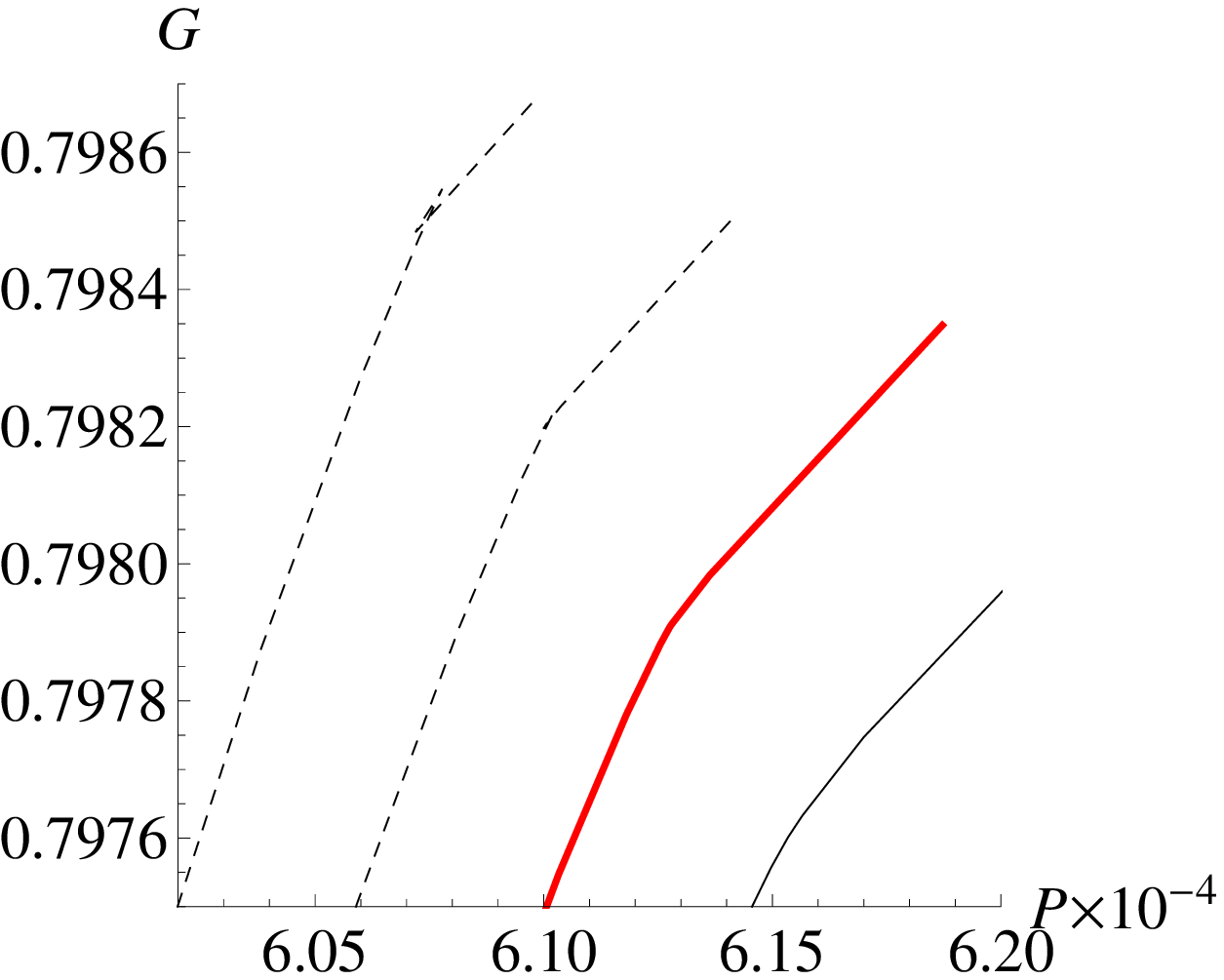}}
\subfigure[~$Q=3$,$\alpha=0.1$] {
\includegraphics[angle=0,width=4cm,keepaspectratio]{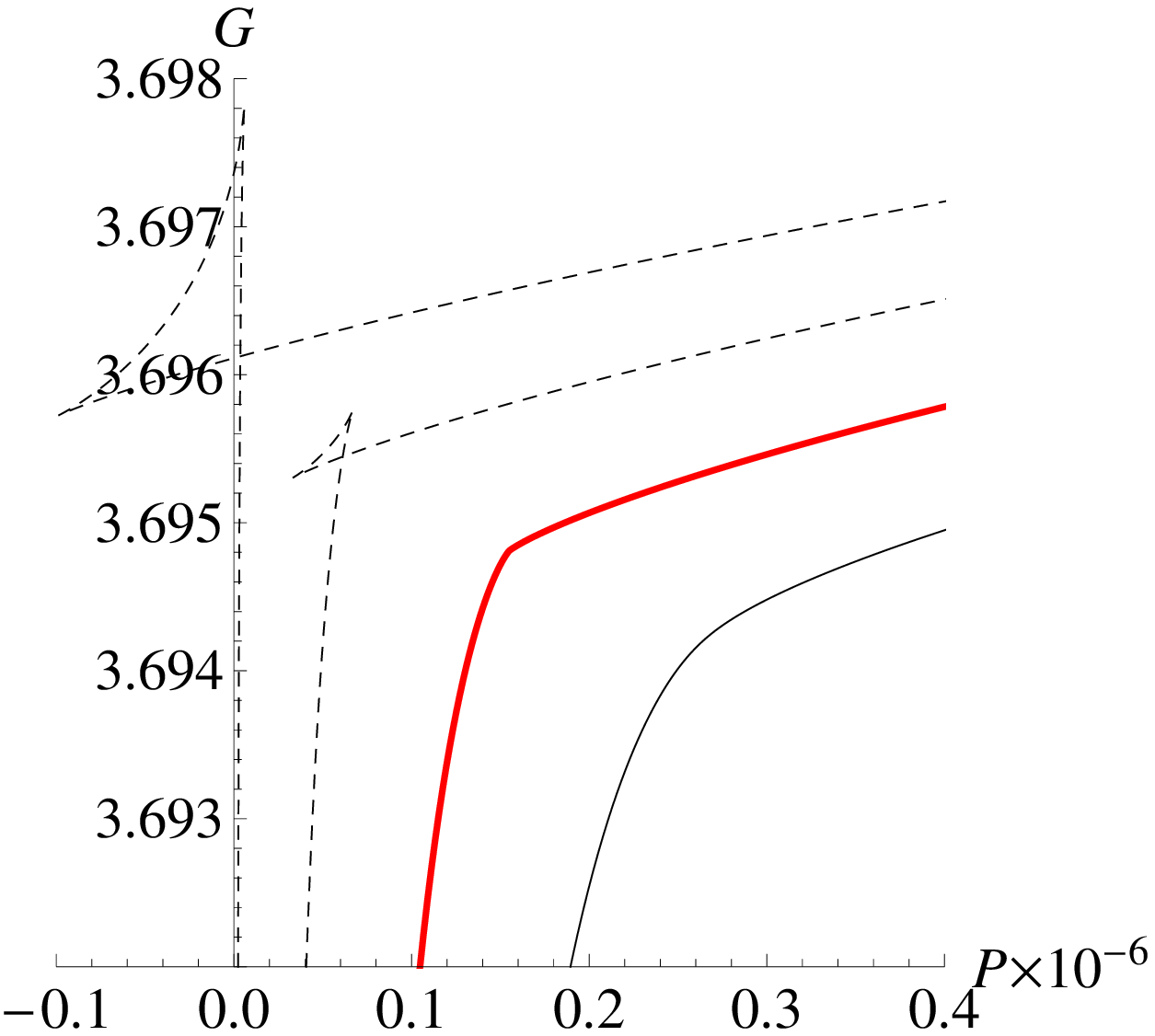}}
\subfigure[~$Q=3$,$\alpha=0.3$] {
\includegraphics[angle=0,width=4cm,keepaspectratio]{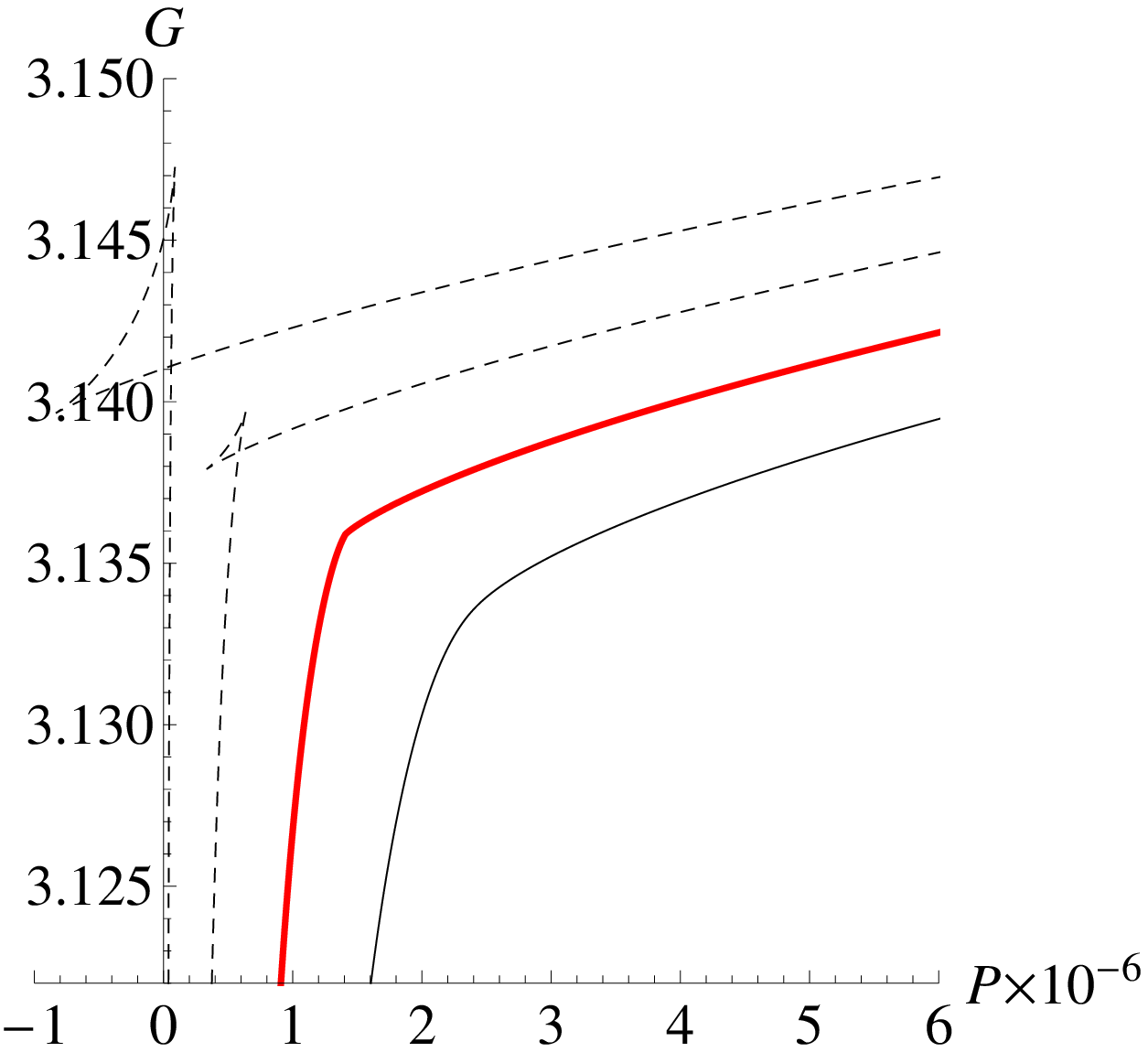}}
\caption[]{\it Effective isotherms in $G-P$ diagrams of $n+1$
dimensional RN-dS-N system as $n=3$. The
dashed lines, the thick solid linesand the thin
solid lines mean the same as that in Fig.1.}} \label{GP}
\end{figure}

In Fig.2,  Gibbs free energy $G$ versus effective pressure $P$ presents a
swallow tail form below the critical temperature. According to Gibbs
free energy criterion, below the critical temperature there exist phase
transitions but above the critical temperature there not.
And the larger $Q$ and $\alpha$, the phase transition is more obvious.
Near the effective critical temperature $T^c$, the phase structure of 
the RN-dS-N system is similar that of van der Waals-Maxwell's liquid-gas system.

\section{Analytical check of the classical Ehrenfest equations at
critical point}

According to Ehrenfest scheme, at the phase transition point, that the chemical
potential (molar Gibbs free energy) and its first partial derivatives are continuous but its second
partial derivatives are mutational indicate that the phase transition
belongs to a second order one. We derive the specific heat $C_P$, expansion
coefficient $\beta$, and the isothermal compressibility $\kappa$ of the RN-dS-N system. 
They are given below as the dimension $n$, electric
charge $q$, and $\alpha $ are regarded as constants.
\begin{equation}
\label{eq25} C_P =T\left( {\frac{\partial S}{\partial T}} \right)_P
=-T\frac{\partial ^2\mbox{G}}{\partial T^2} =T \left( {\frac{\left(
{\frac{\partial S}{\partial x}} \right)_{r_c } \left(
{\frac{\partial P}{\partial r_c }} \right)_x -\left( {\frac{\partial
S}{\partial r_c }} \right)_x \left( {\frac{\partial P}{\partial x}}
\right)_{r_c } }{\left( {\frac{\partial T}{\partial x}} \right)_{r_c
} \left( {\frac{\partial P}{\partial r_c }} \right)_x -\left(
{\frac{\partial T}{\partial r_c }} \right)_x \left( {\frac{\partial
P}{\partial x}} \right)_{r_c } }} \right),
\end{equation}
\begin{equation}
\label{eq26} \beta =\frac{1}{V}\left( {\frac{\partial V}{\partial
T}} \right)_P =\frac{1}{V}\frac{\partial ^2\mu }{\partial T\partial
P} =\frac{1}{V}\left( {\frac{\left( {\frac{\partial V}{\partial x}}
\right)_{r_c } \left( {\frac{\partial P}{\partial r_c }} \right)_x
-\left( {\frac{\partial V}{\partial r_c }} \right)_x \left(
{\frac{\partial P}{\partial x}} \right)_{r_c } }{\left(
{\frac{\partial T}{\partial x}} \right)_{r_c } \left(
{\frac{\partial P}{\partial r_c }} \right)_x -\left( {\frac{\partial
T}{\partial r_c }} \right)_x \left( {\frac{\partial P}{\partial x}}
\right)_{r_c } }} \right),
\end{equation}
\begin{equation}
\label{eq27} \kappa =-\frac{1}{V}\left( {\frac{\partial V}{\partial
P}} \right)_T =-\frac{1}{V}\frac{\partial ^2\mu }{\partial P^2}
=-\frac{1}{V}\left( {\frac{\left( {\frac{\partial V}{\partial x}}
\right)_{r_c } \left( {\frac{\partial T}{\partial r_c }} \right)_x
-\left( {\frac{\partial V}{\partial r_c }} \right)_x \left(
{\frac{\partial T}{\partial x}} \right)_{r_c } }{\left(
{\frac{\partial P}{\partial r_c }} \right)_x \left( {\frac{\partial
T}{\partial x}} \right)_{r_c } -\left( {\frac{\partial P}{\partial
x}} \right)_{r_c } \left( {\frac{\partial T}{\partial r_c }}
\right)_x }} \right).
\end{equation}
In Fig.3, $C_P-P$, $\beta-P$, and $\kappa-P$ curves for given values of
$Q$ and $\alpha$ are shown.  According to a preliminary inspection
the phase transitions at critical temperature belong to a second order one.
The phase transitions below critical temperature are more complex.

\begin{figure}[!htbp]
\center{\subfigure[~$Q=1$,$\alpha=0.1$] {
\includegraphics[angle=0,width=4cm,keepaspectratio]{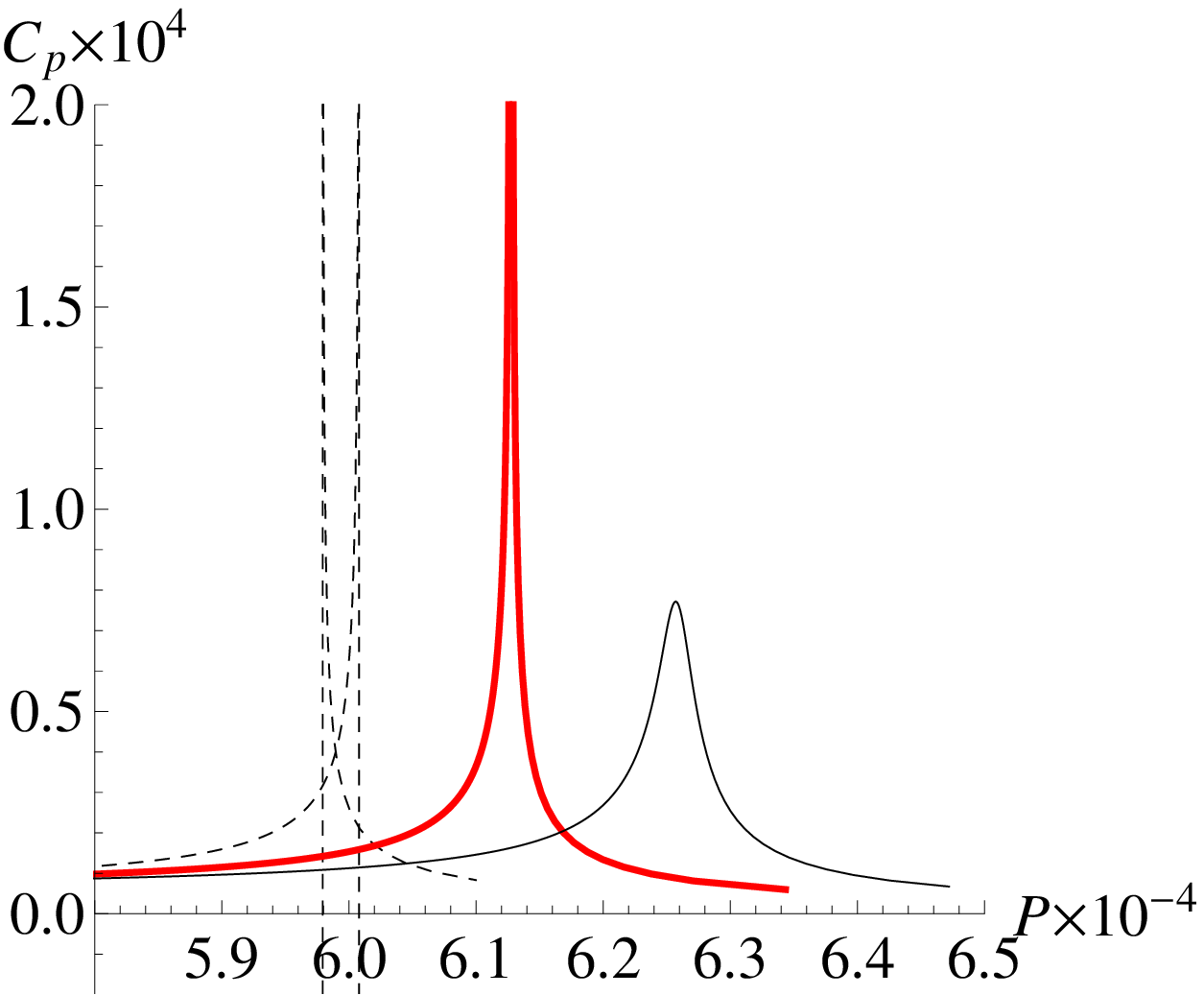}}
\subfigure[~$Q=1$,$\alpha=0.1$] {
\includegraphics[angle=0,width=4cm,keepaspectratio]{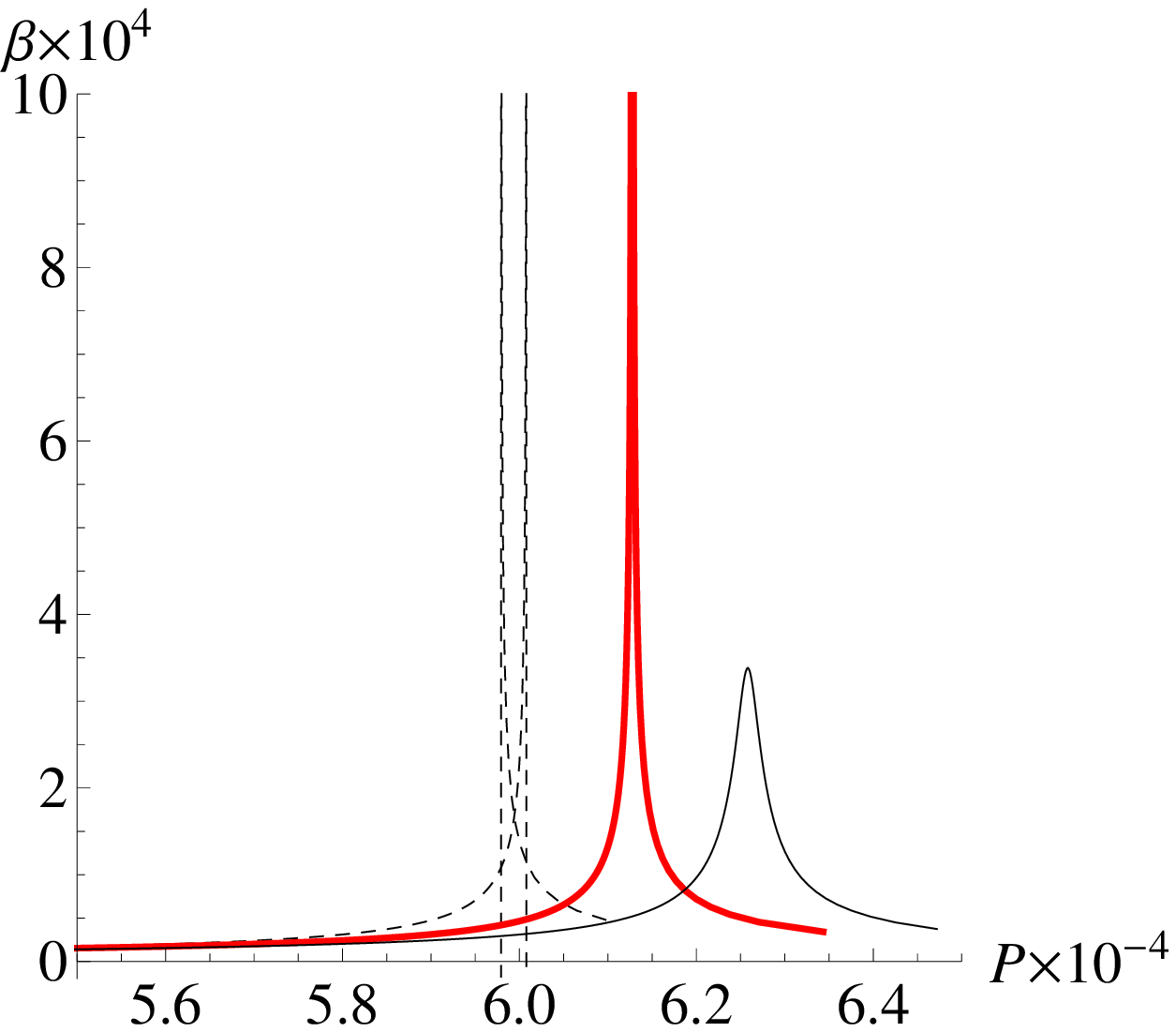}}
\subfigure[~$Q=1$,$\alpha=0.1$] {
\includegraphics[angle=0,width=4cm,keepaspectratio]{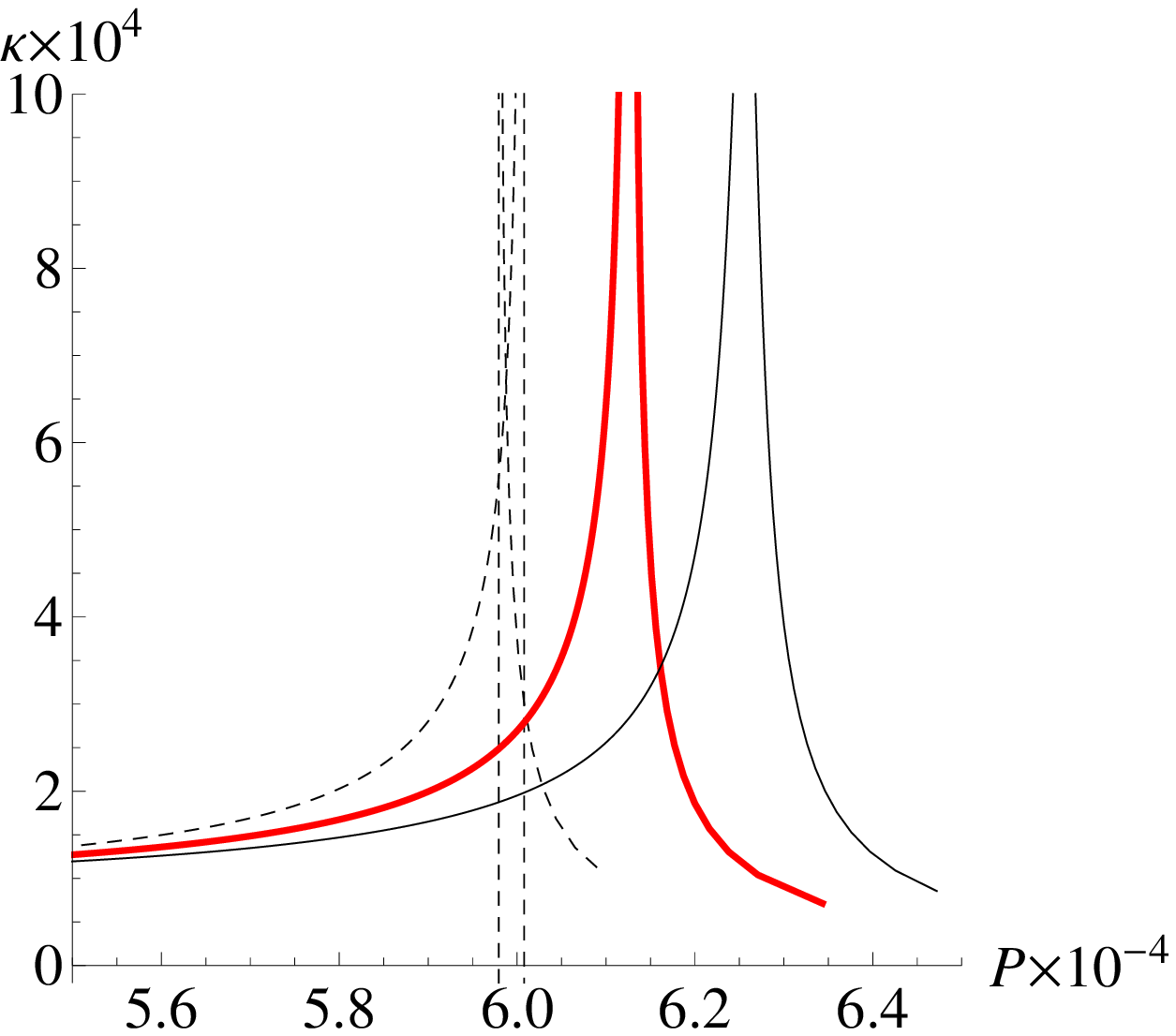}}
\subfigure[~$Q=3$,$\alpha=0.3$] {
\includegraphics[angle=0,width=4cm,keepaspectratio]{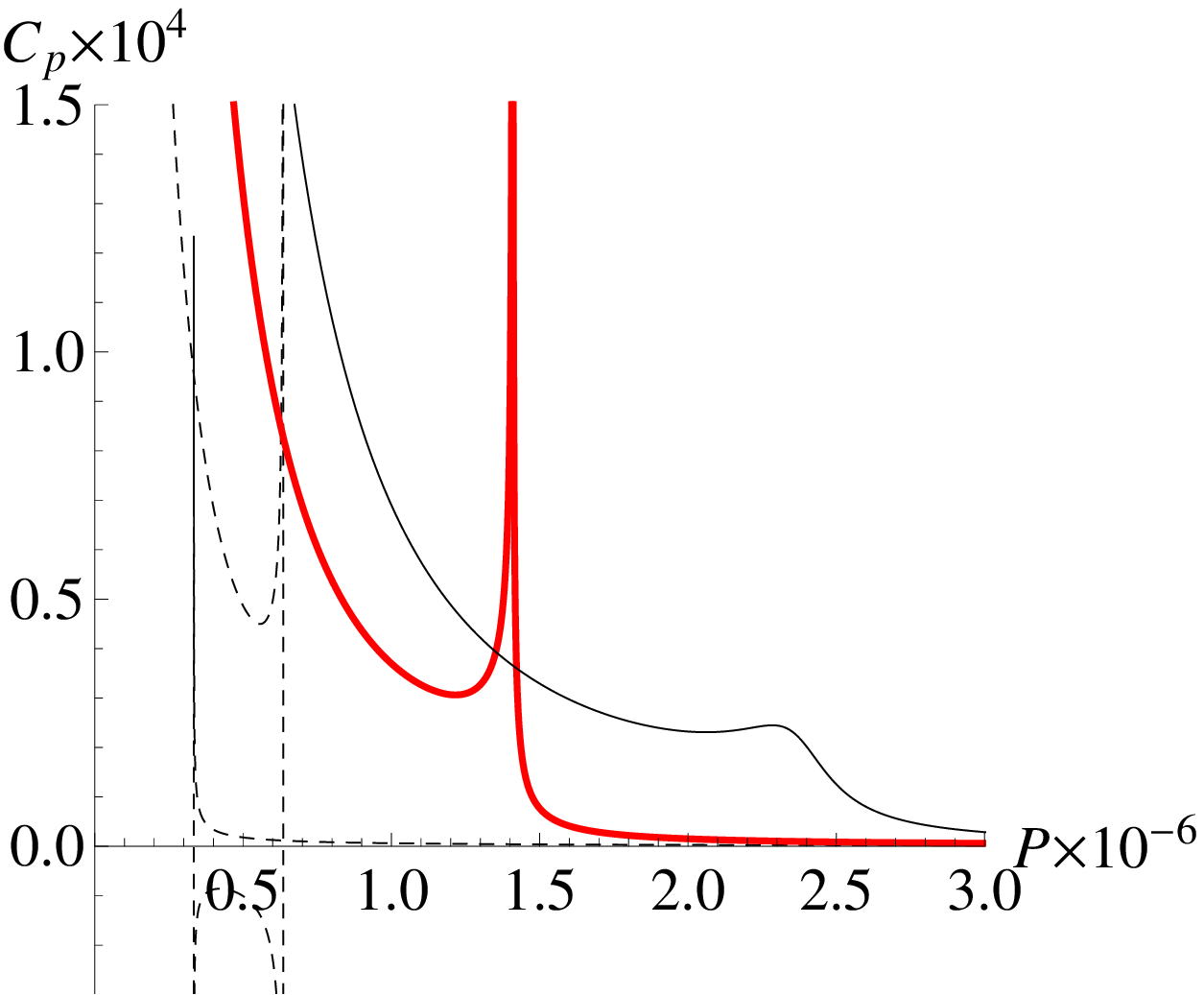}}
\subfigure[~$Q=3$,$\alpha=0.3$] {
\includegraphics[angle=0,width=4cm,keepaspectratio]{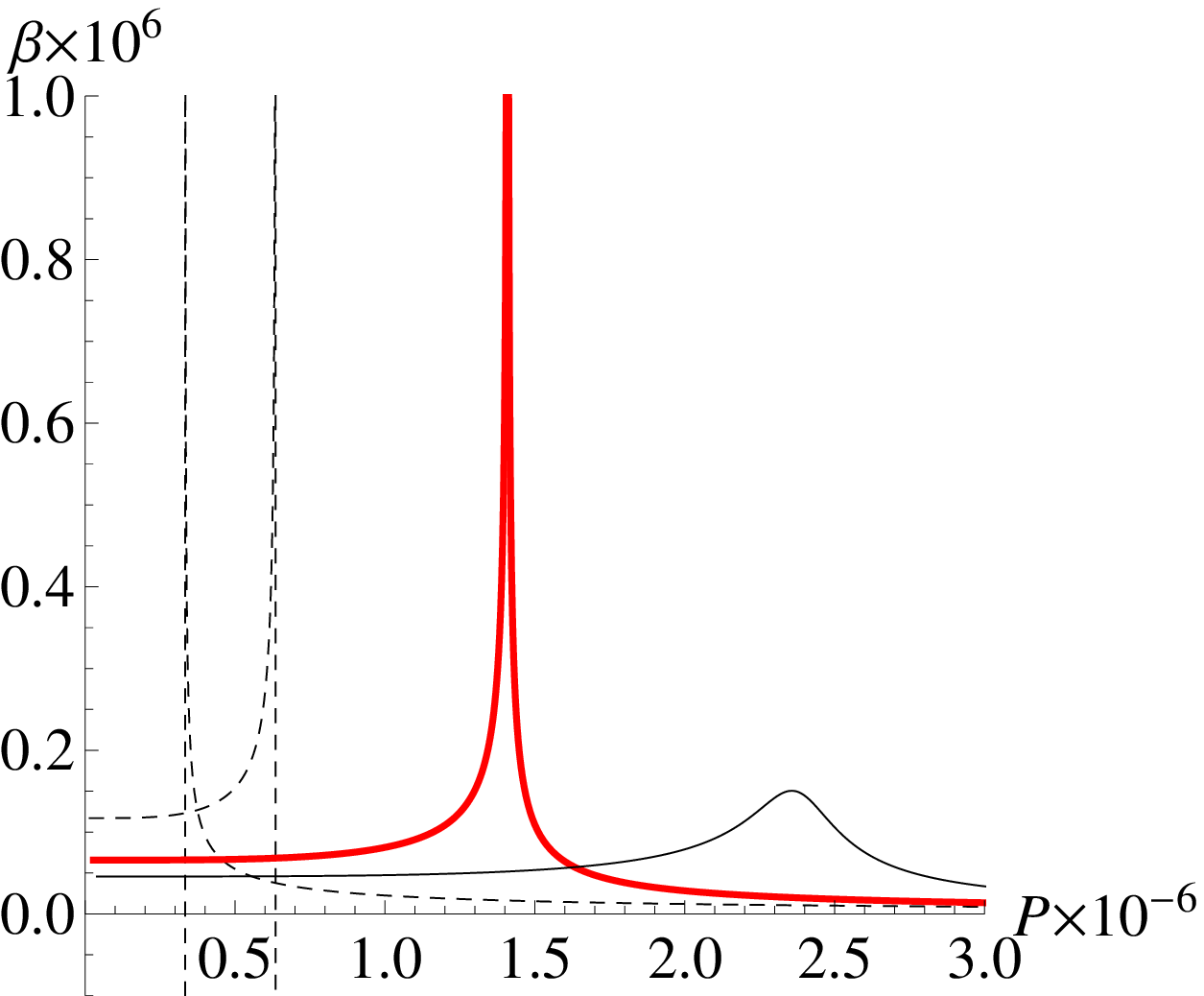}}
\subfigure[~$Q=3$,$\alpha=0.3$] {
\includegraphics[angle=0,width=4cm,keepaspectratio]{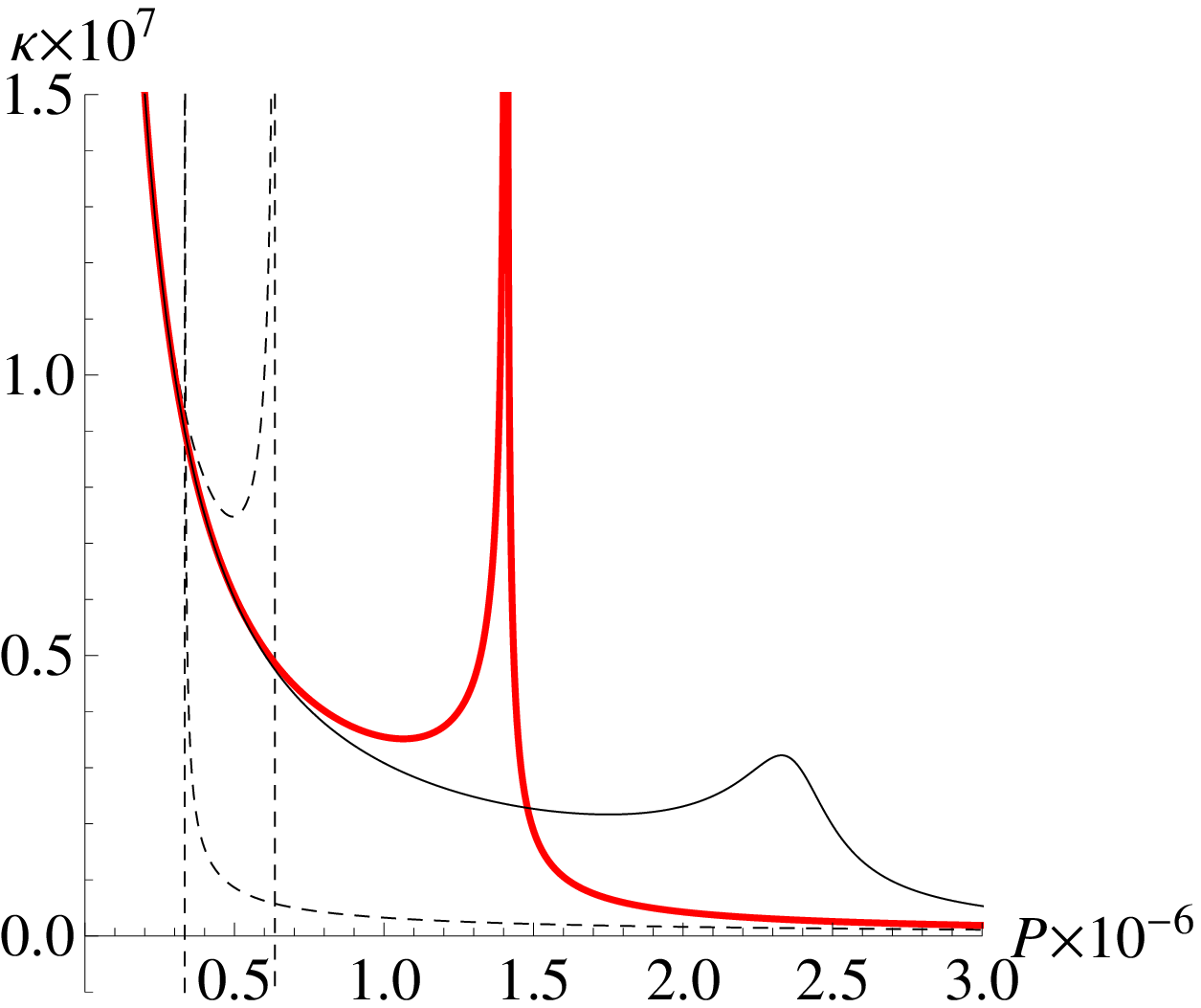}}
\caption[]{\it Effective isotherms in $C_P-P$, $\beta-P$, $\kappa-P$
diagrams of $n+1$
dimensional RN-dS-N system as $n=3$.The dashed lines, thick solid lines, and thin solid lines
represent the same as that in Fig. 1.}}\label{CpBetaKappa}
\end{figure}

The Ehrenfest Equtions
\begin{equation}
\label{eq28} \left( {\frac{\partial P}{\partial T}} \right)_S
=\frac{C_P^2 -C_P^1 }{T^cV^c(\beta _q^2 -\beta _q^1 )}=\frac{\Delta
C_P }{T^cV^c\Delta \beta _q },
\end{equation}
\begin{equation}
\label{eq29} \left( {\frac{\partial P}{\partial T}} \right)_V
=\frac{\beta _q^2 -\beta _q^1 }{\kappa_q^2 -\kappa_q^1 }=\frac{\Delta \beta _q
}{\Delta \kappa_q }.
\end{equation}
The superscripts 1 and 2 represent phases 1 and 2 respectively,
and the subscript $q$ represents $q$ remains unchanged.

According to Maxwell relations
\begin{equation}
\label{eq30} \left( {\frac{\partial V}{\partial S}} \right)_P
=\left( {\frac{\partial T}{\partial P}} \right)_S , \quad \left(
{\frac{\partial V}{\partial S}} \right)_T =\left( {\frac{\partial
T}{\partial P}} \right)_V \quad,
\end{equation}
we can get
\begin{equation}
\label{eq31} \left[ {\left( {\frac{\partial P}{\partial T}}
\right)_S } \right]^c=\left[ {\left( {\frac{\partial S}{\partial V}}
\right)_P } \right]^c, \quad \left[ {\left( {\frac{\partial
S}{\partial V}} \right)_T } \right]^c=\left[ {\left( {\frac{\partial
P}{\partial T}} \right)_V } \right]^c,\quad .
\end{equation}
Note that the footnote `c' denotes the values of physical quantities
at the critical point. Substitute Eq.(\ref{eq31}) into
Eqs. (\ref{eq28}) and (\ref{eq29}),

\begin{equation}
\label{eq32} \frac{\Delta C_P }{T^cV^c\Delta \beta _q }=\left[
{\left( {\frac{\partial S}{\partial V}} \right)_P } \right]^c, \quad
\frac{\Delta \beta _q }{\Delta \kappa_q }=\left[ {\left( {\frac{\partial
S}{\partial V}} \right)_T } \right]^c\quad.
\end{equation}

From Eq.(\ref{eq22}), at the critical point, there is

\begin{equation}
\label{eq33} \frac{\partial (P,T)}{\partial (r_c ,x)}=0.
\end{equation}
Moreover

\begin{equation}
\label{eq34+} \left( {\frac{\partial S}{\partial V}} \right)_P
=\frac{\frac{\partial (S,P)}{\partial (r_c ,x)}}{\frac{\partial
(V,P)}{\partial (r_c ,x)}},  \quad \left( {\frac{\partial
S}{\partial V}} \right)_T =\frac{\frac{\partial (S,T)}{\partial (r_c
,x)}}{\frac{\partial (V,T)}{\partial (r_c ,x)}}.
\end{equation}

Substitute (\ref{eq33}) into(\ref{eq34+}), one can get
\begin{equation}
\label{eq34}
\left( {\frac{\partial S}{\partial V}} \right)_P^c =\left( {\frac{\partial
S}{\partial V}} \right)_T^c .
\end{equation}

So far, we have proved the validity of both Ehrenfest equations 
at the critical point. Utilizing Eqs. (\ref{eq34}) and (\ref{eq32}),
the Prigogine-Defay (PD) ratio can be calculated,
\begin{equation}
\label{eq35}
\Pi =\frac{\Delta C_P \Delta \kappa_q }{T^cV^c\Delta \beta _q^2 }=1.
\end{equation}

Hence the phase transition occurring at $T=T^c$ is a second order
equilibrium transition. The conclusion is similar to that in AdS spacetime\cite{Mo2014}.

\section{Discussions and Conclusions}

In this paper, considering the correlation of the thermodynamic
quantities corresponding to the two horizons respectively in dS
spacetime, we discuss thermodynamic properties of RN-dS spacetime
with nonlinear source by a set of effective thermodynamic
quantities, which reflect thermodynamic property of the two horizons
and the whole dS spacetime as a thermodynamic system.

we investigate the phase transition of the RN-dS-N system, and found that the nonlinearity parameters $\alpha$
along with electric charge influence the phase structure of the
system, which can be seen in Table 1, Fig.1 and Fig.2. When the
effective temperature is below critical temperature, a phase
transition can happen, which can be seen in Fig. 2 and inferred
from Gibbs free energy criterion.

We carry out an analytical check of Ehrenfest equations and derive
the specical heat, expansion coefficient, and the isothermal
compressibility of the effective RN-dS-N system. And find that the
RN-dS-N system undergoes a second order
equilibrium phase transition at the critical point. This result is similar to
the nature of Van der Waala liquid-gas phase transition at the critical point.

\begin{acknowledgments}\vskip -4mm

This work is supported by NSFC under Grant
No.11475108, by the doctoral Sustentation Fund of Shanxi Datong University (2015-B-10), 
and by the Natural Science Foundation for
Young Scientists of Shanxi Province,China (Grant No.2012021003-4).

\end{acknowledgments}

\end{document}